\documentclass[12pt]{article}
\usepackage{graphicx}
\def \beq{\begin{equation}}
\def \eeq{\end{equation}}

\begin{document}

\centerline{\bf HEAVY-QUARK EXOTICS}
\medskip

\centerline{\it Jonathan L. Rosner}
\centerline{\it Enrico Fermi Institute, University of Chicago}
\centerline{\it 5640 Ellis Avenue, Chicago, IL 60637}
\medskip

\centerline{Talk presented at Workshop on Electroweak Processes of Hadrons}
\centerline{Bled, Slovenia, July 15--19, 2019}

\begin{quote}

The heavy quarks $c$ and $b$ stabilize exotic meson $(qq\bar q \bar q)$
and baryon $(qqqq \bar q)$ states. We discuss work with M. Karliner on
molecules containing $c \bar c$ and $b \bar b$; the first doubly charmed baryon;
isospin splittings; $\Xi_{cc}^+ = ccd$ and $\Omega_{cc} = ccs$ masses;
lifetimes; tetraquarks stable under strong and electromagnetic decay; excited
$\Omega_{c}$ states; and P-wave excitation energies.
\end{quote}

In 1964 M. Gell-Mann \cite{GM} and G. Zweig \cite{GZ} proposed that
the known mesons were $q \bar q$ and baryons $qqq$, with quarks known at the
time $u$ (``up''), $d$ (``down''), and $s$ (``strange'') having charges
(2/3,--1/3,--1/3).  Mesons and baryons would then have integral charges.
Mesons such as $qq \bar q \bar q$ and baryons such as $qqqq \bar q$
would also have integral charges. Why weren't they seen?
They {\it have} now been seen, as ``molecules'' of heavy-quark hadrons or as
deeply bound states involving heavy quarks (charm and bottom).

An early prediction of exotics was based on duality between $s$-channel and
$t$-channel processes \cite{JLR68}.  In antiproton-proton scattering,
$q \bar q$ is dual to $q q \bar q \bar q$, predicting ``exotic'' $q q \bar q
\bar q$ mesons.  Where would they occur?  One picture of resonance formation
is based on $q \bar q$ annihilation \cite{JLR72}.  If $p^*$ is the momentum
of each colliding particle in their center of mass, the first (meson-meson,
meson-baryon) resonance forms for $p^* < (350,250)$ MeV.  Optical reasoning
then leads one to expect the first baryon-antibaryon resonance to form for
$p^* < 200$ MeV.  The first ``baryonium'' candidate was actually the pion
\cite{FY49}, envisioned as a nucleon-antinucleon bound state.

A QCD string picture can distinguish a standard $q \bar q$ meson, a standard
baryon, and an exotic meson from one another.  If decays occur via quark pair
production (breaking of a QCD string), a $qq \bar q \bar q$ meson will either 
decay to baryon-antibaryon or to an ordinary meson plus an exotic one.  It was
proposed \cite{JLR68} to search for exotic mesons in the backward direction
of a meson-baryon collision.  Such exotics may fall apart into meson pairs and
may be too broad to show up as distinct resonant peaks.  No resonances made of
$u,d,s$ have been seen which would correspond to $q q \bar q \bar q$ but not
$q \bar q$ (e.g., $u u \bar d \bar s$ decaying to $K^+ \pi^+$).  Similarly,
pentaquark states ($4q \bar q$) made only of $u,d,s$ have not been confirmed.
R. Jaffe made an extensive study of $qq \bar q \bar q$ states within the bag
model of QCD \cite{RJ}.  Light diquark-antidiquark states could be familiar
ones with masses of a GeV or less.

The situation changed with heavy quarks $c$ (charm) and $b$, which act to
stabilize exotic configurations.  The charmed quark was introduced in
1964 to preserve lepton-quark symmetry \cite{BGetal}.  The suppression of
higher-order weak corrections led Glashow, Iliopoulos, and Maiani \cite{GIM}
to estimate $m_c \simeq 2$ GeV/$c^2$, while Gaillard and Lee (1973) \cite{GL}
studied the charmed quark's role in gauge theories.  Evidence for the charmed
quark $c$ appeared in the $c \bar c$ bound state $J/\psi$ \cite{Ting,SLAC}.  An
abundant charmonium ($c \bar c$) spectrum is still evolving.

Particles with one charmed quark also display a rich spectrum.  The large
value of $m_c$ allows nonrelativistic quantum mechanics to provide some
insights.  Evidence for a third quark-lepton family began with observation of
the $\tau$ lepton \cite{tau}.  The quark-lepton analogy then implied the
existence of a quark doublet ($t$ [top], $b$ [bottom]), first predicted by
Kobayashi and Maskawa \cite{KM} to explain CP violation.  Evidence for the $b$
quark came from observation in 1977 at Fermilab of the first members of the
$\Upsilon$ family of spin--1 $b \bar b$ particles produced in proton-proton
interactions, decaying to $\mu^+ \mu^-$ \cite{E288}. Today there is
a rich spectroscopy both of $b \bar b$ states and of ``$B$'' mesons containing
a single $b$ quark.  Decays of particles with $b$ quarks are an active field.
The top quark, discovered in 1995 at the Fermilab Tevatron \cite{top}, has a
mass $m_t \simeq 173$ GeV so large that it decays too rapidly to have
interesting spectroscopy.

The first genuine exotic, $X(3872)$, was seen decaying to $J/\psi
\pi^+ \pi^-$ by the Belle Collaboration in 2003 \cite{3872}, and confirmed by
CDF \cite{CDFX}, D0 \cite{D0X}, and BaBar \cite{BaBarX}.  Its identification as
a $D^0 \bar D^{*0} + {\rm c.c.}$ molecule comes from its proximity to
$D^0 \bar D^{*0}$ threshold: $M(X) = (3871.69 \pm 0.17)$ MeV $\simeq M(D^0) +
M(\bar D^{*0}) = (3871.68 \pm 0.07)$ MeV.  Its decay $X \to \gamma J/\psi$ is
seen, implying $C(X) = +$ and some admixture of $c \bar c$ in its wave
function.  The angular distribution of its decay products implies $J^{PC} =
1^{++}$ as expected for an S-wave state of $D^0 \bar D^{*0}+{\rm c.c.}$
\cite{Xangdist}. $C$ invariance implies the $\pi^+ \pi^-$ pair in its decay
has negative $C$, as in a $\rho$ meson.  The large value of $M(D^{(*)+})-%
M(D^{(*)0})$ implies little $D^{(*)\pm}$ in its wave function.  The comparable
rates for $\Gamma(X \to \omega J/\psi)$ and $\Gamma(X \to J/\psi \rho)$ are
what one would expect for a state with a $c \bar c u \bar u$ admixture.  
In addition to the $X(3872)$ (a mixture of $2^3P_1~c \bar c$ and $J^{PC}
= 1^{++}~c \bar c u \bar u$) one expects an orthogonal mixture, typically
above 3900 MeV in potential models.

\begin{figure}[h]
\includegraphics[width=0.32\textwidth]{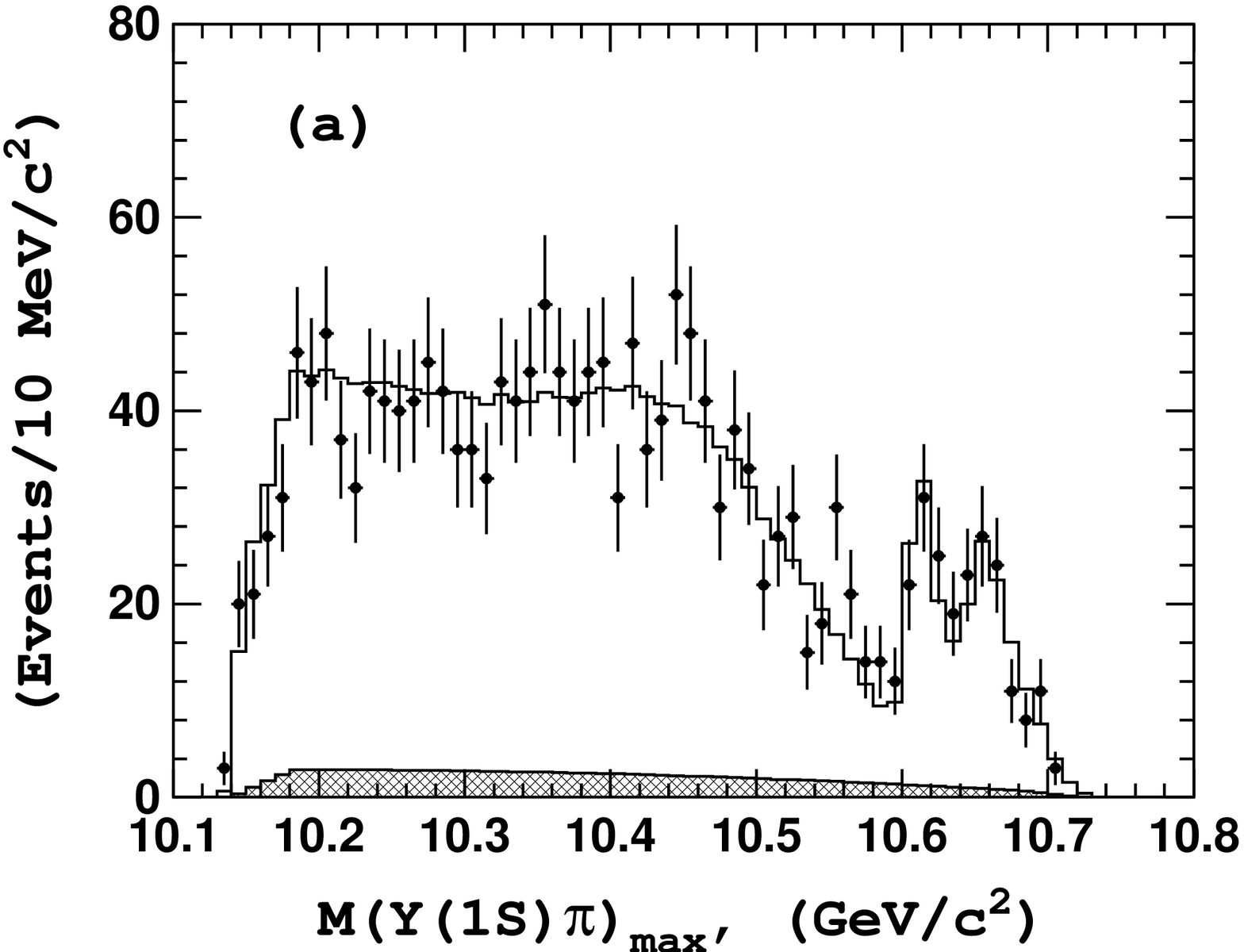}
\includegraphics[width=0.32\textwidth]{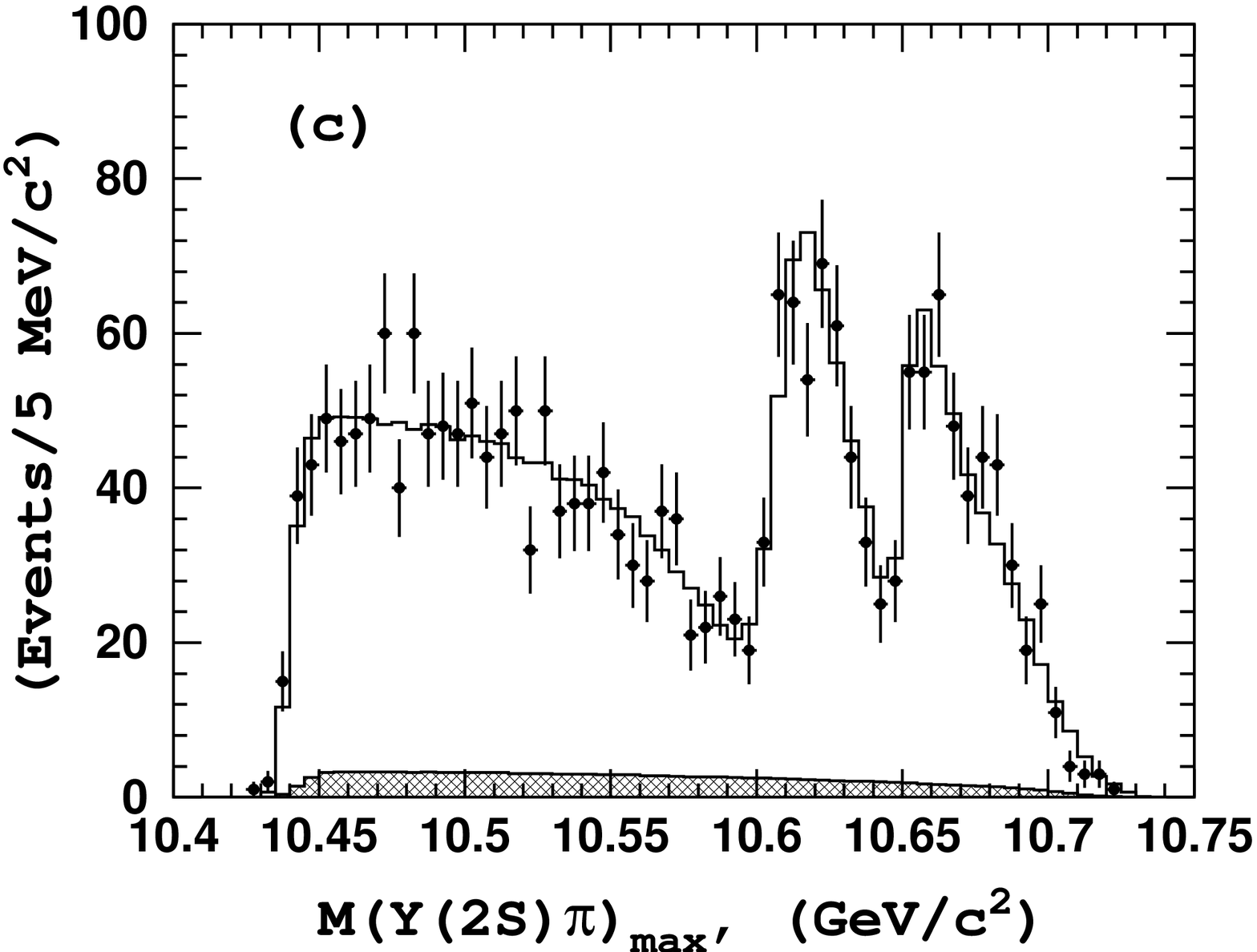}
\includegraphics[width=0.32\textwidth]{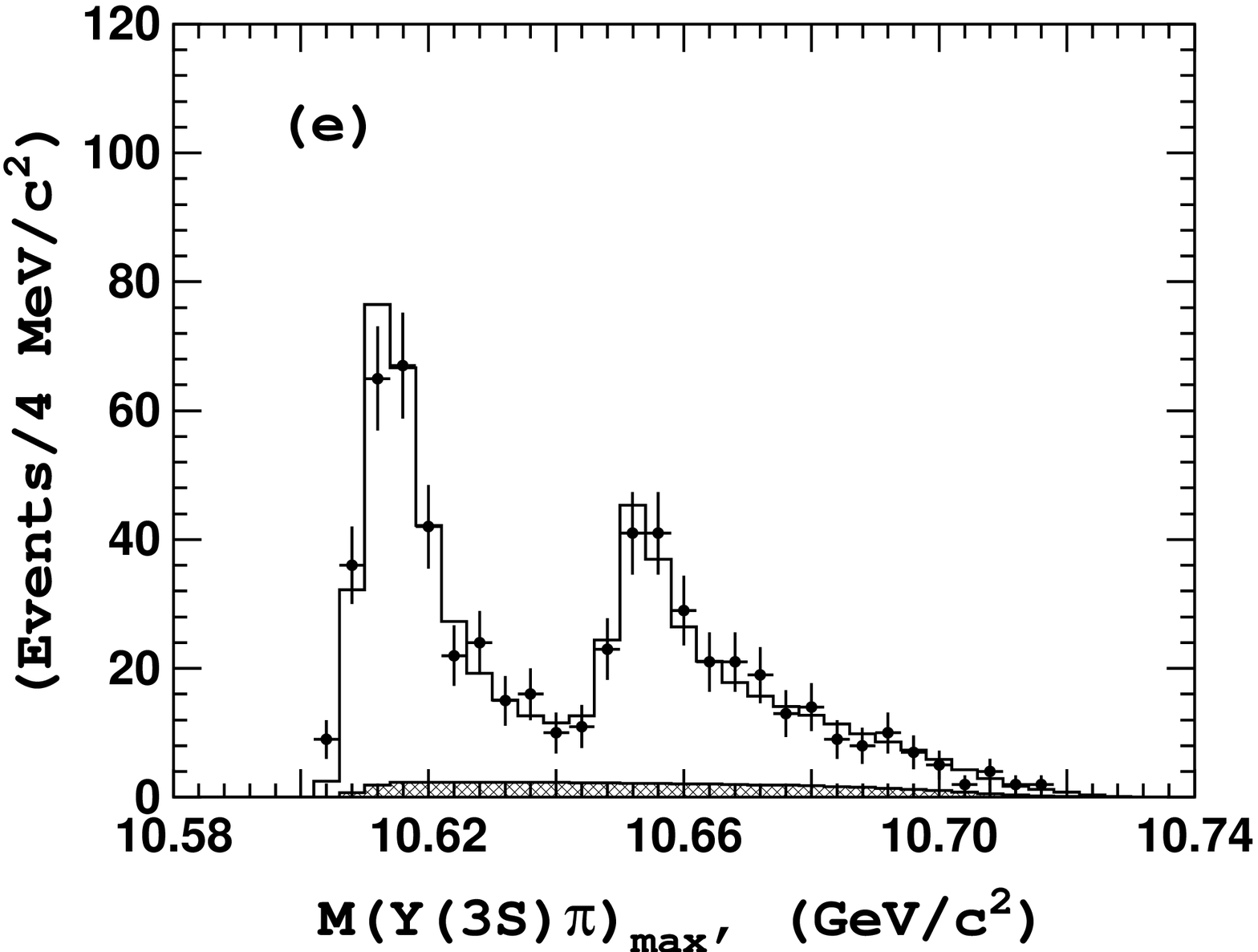}
\caption{Mass spectra $M(\Upsilon(1S,2S,3S) \pi^+)$ in $\Upsilon(10865) \to
\Upsilon(1S,2S,3S) \pi^+ \pi^-$ \cite{BelleZ}.
\label{fig:zbs}}
\end{figure}

The Belle Collaboration saw unexpected structures $Z_b(10610,10650)$
\\ in $M[\pi^\pm \Upsilon(1S,2S,3S)]$ when studying $\Upsilon(10865) \to
\Upsilon(1S,2S,3S) \pi^+ \pi^-$ \cite{BelleZ} (Fig.\ \ref{fig:zbs}).
All spectra showed peaks at $M(\Upsilon(nS) \pi = 10.61$ and $10.65$ GeV,
within a few MeV of $M(B) + M(\bar B^*)$ and $M(B^*) + M(\bar B^*)$.  These
look like S-wave molecules of $B \bar B^* (+ {\rm c.c.})$ and $B^* \bar B^*$.

\begin{figure}[h]
\hskip 0.4in
\includegraphics[width=0.88\textwidth]{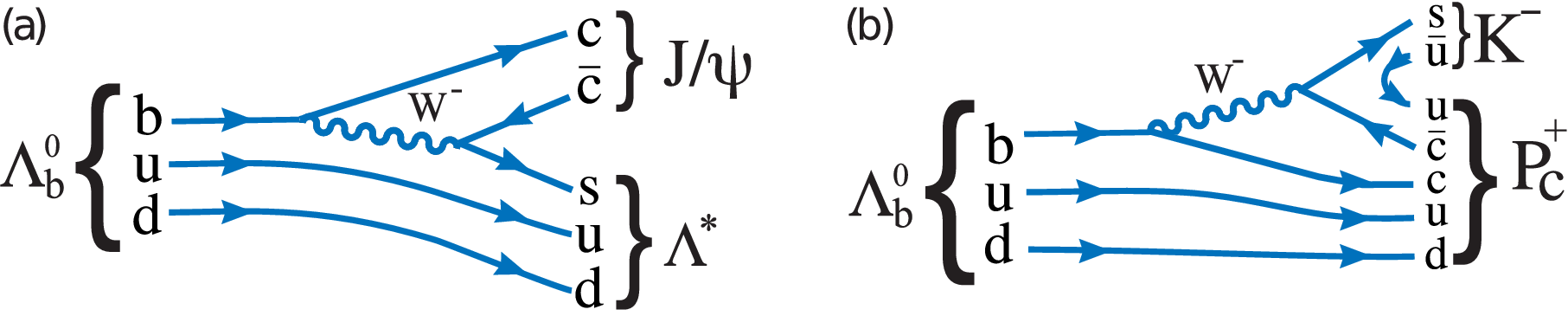}
\caption{Production mechanisms in $\Lambda_b$ decays. Left: $\Lambda^*$
excitation; right: $P_c$ excitation.
\label{fig:pcprod}}
\end{figure}

Evidence for $c \bar c uud$ configurations has been provided by LHCb
\cite{LHCb1}, who observed bumps in the $J/\psi~p$ invariant mass in the decay
$\Lambda_b \to K^- J/\psi~p$ at 4380 and 4450 MeV.  (See Fig.\ \ref{fig:pcprod}
for a production mechanism.)

\begin{figure}
\includegraphics[width=0.9\textwidth]{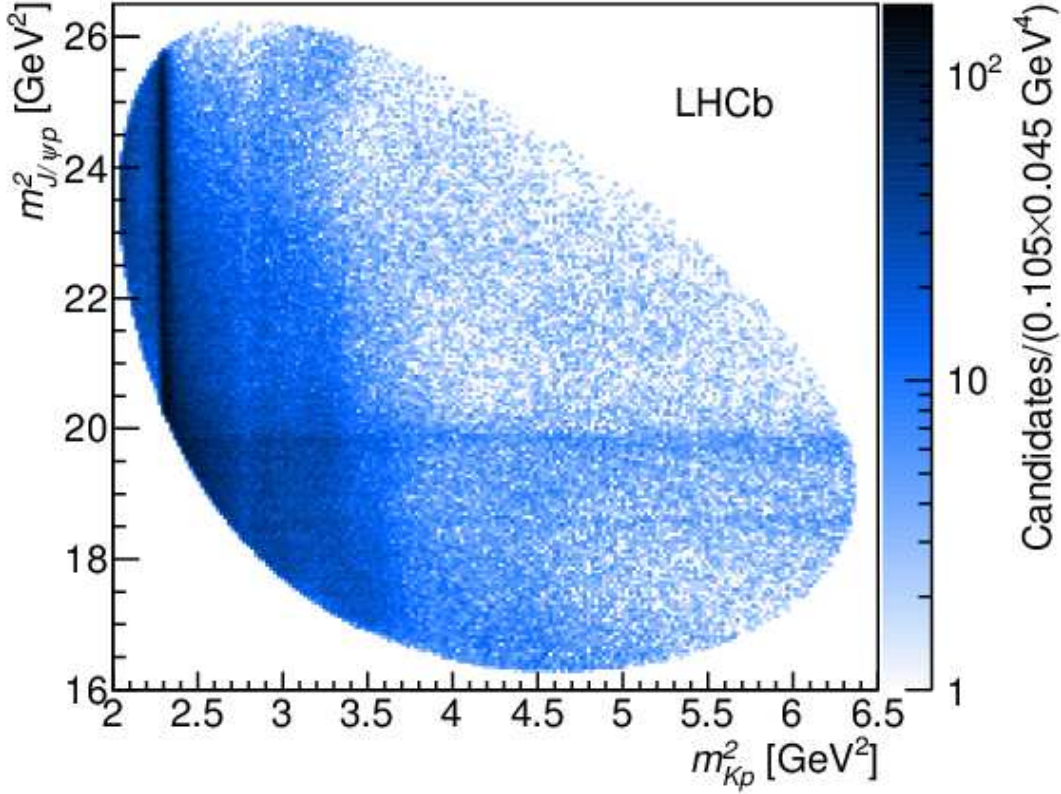}
\caption{$K^- J/\psi~p$ Dalitz plot in $\Lambda_b \to K^- J/\psi~p$
\cite{LHCb2}
\label{fig:pcdp}}
\end{figure}

The $K^- J/\psi~p$ Dalitz plot (Fig.\ \ref{fig:pcdp}) is populated by many
$I=0~K^-p$ states.  In an updated result \cite{LHCb2}, LHCb
sees {\it three} narrow $J/\psi$ resonances at 4311.9, 4440.3, 4457.3 MeV,
with widths 9.8, 20.6, 6.4 MeV.  The masses are near $\Sigma_c \bar D$ and
$\Sigma_c \bar D^*$ thresholds; if these are molecules, their binding mechanism
is unclear.  One-pion exchange can't couple to $D \bar D$; $\pi^+\pi^-$
exchange may favor $\Sigma_c \bar D$ over $D \bar D$: the lowest intermediate
state is $\Lambda_c \bar D^*$ vs.\ $D^* \bar D^*$.
The asymmetric behavior along $M(J/\psi~p)$ bands indicates
interference with opposite-parity amplitude(s).

\begin{figure}[h]
\includegraphics[width=0.48\textwidth]{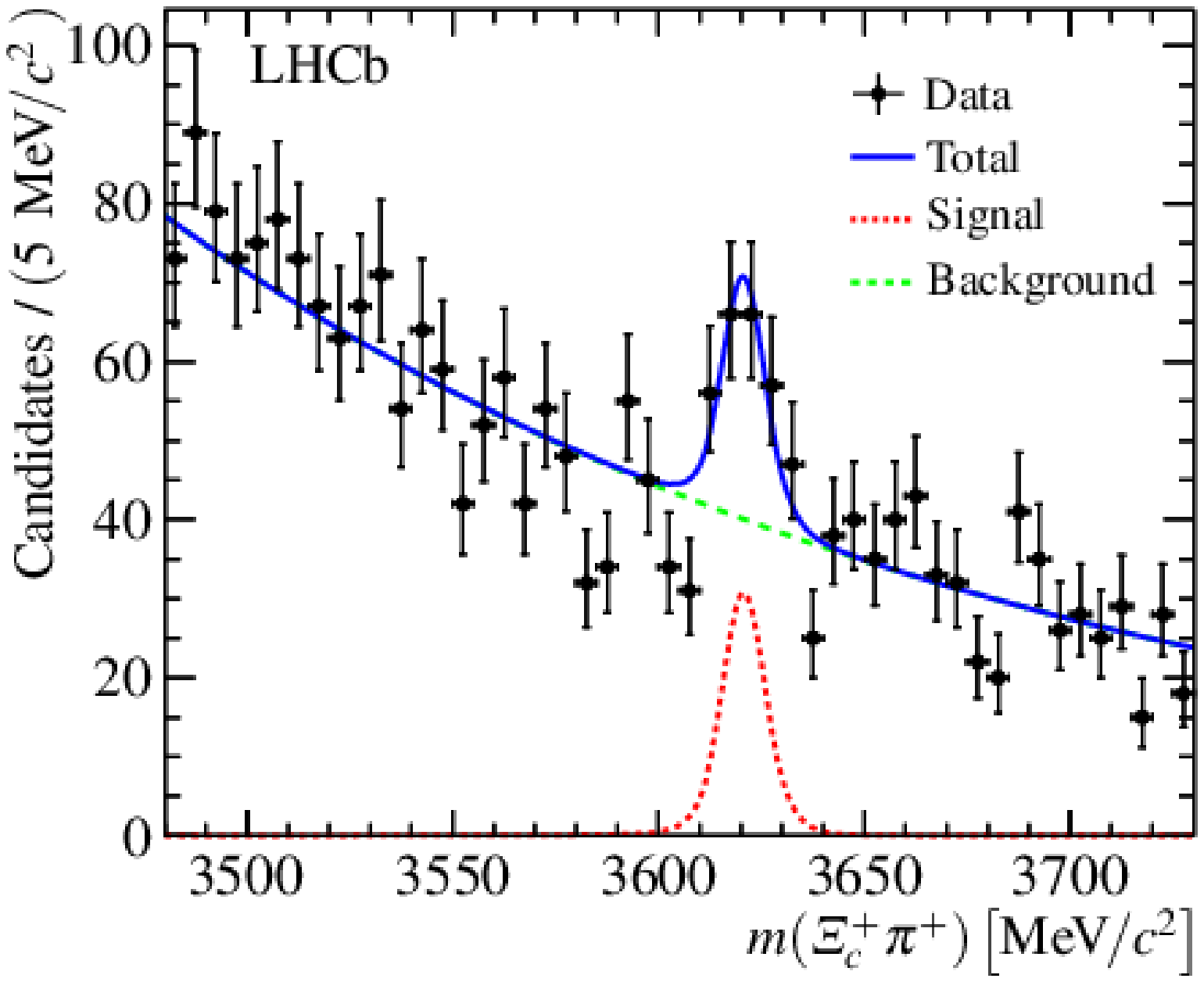}
\includegraphics[width=0.48\textwidth]{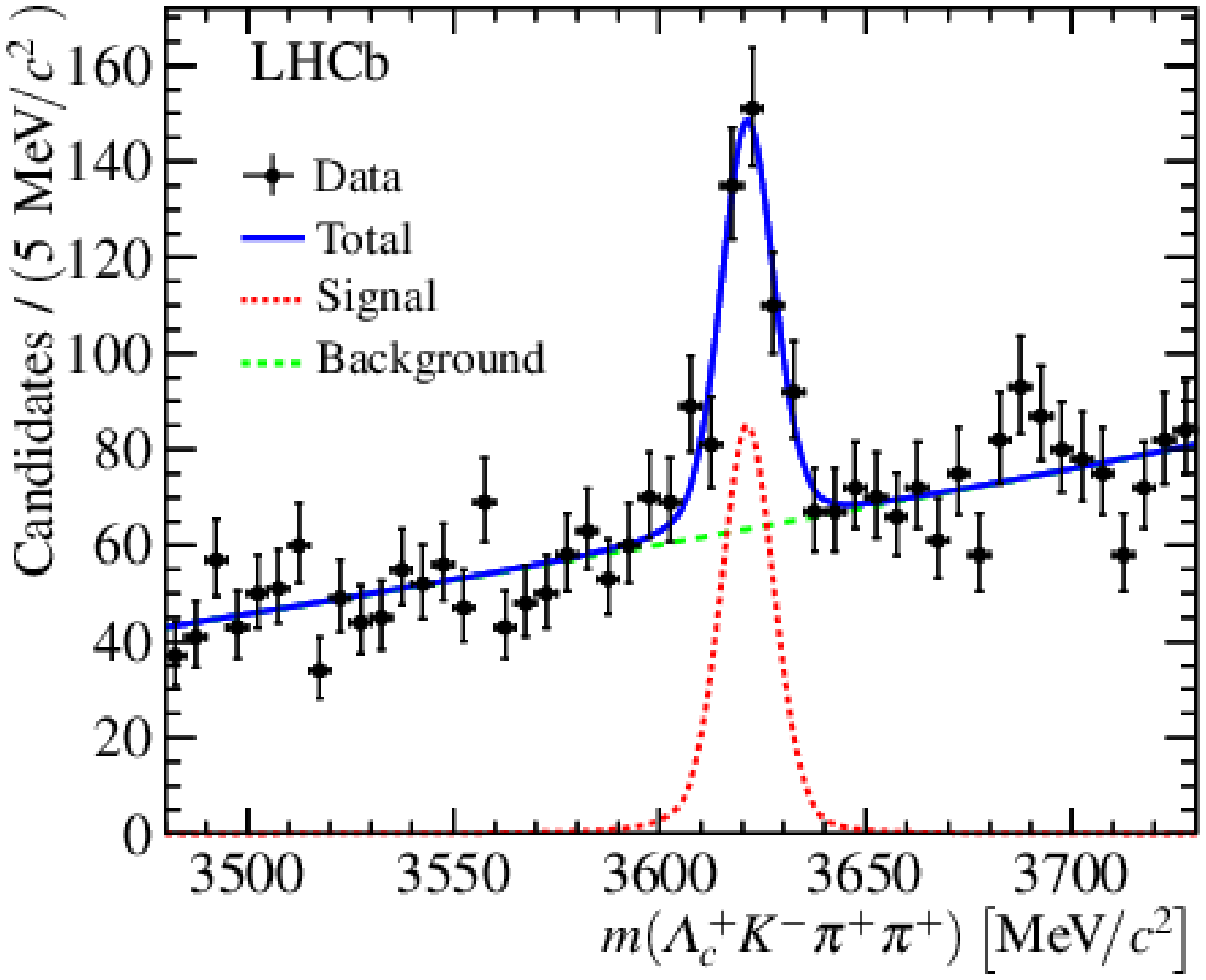}
\caption{Spectra with evidence for $\Xi_{cc}^{++}$.
\label{fig:spec}}.
\end{figure}

So far we have discussed $Q \bar Q q \bar q'$ or $Q \bar Q qqq'$ states, where
$Q$ = heavy, $q,q'$ = light.  Can we predict masses of (simpler) $QQ'q$
systems?  The SELEX Collaboration at Fermilab \cite{SELEX} claimed states
$\Xi_{cc}^{++}(3520) = ccu$ and $\Xi_{cc}^+(3460) = ccd$ which were not
confirmed by others.  Using constituent-quark masses, hyperfine splittings,
and estimates of $QQ'$ binding M. Karliner and I \cite{PRD90}
predicted the masses in Table 1.  In 2017 the LHCb Collaboration found a
$\Xi_{cc}^{++}$ candidate with mass $3621.40 \pm 0.78$ MeV \cite{LHCXicc},
in accord with our estimate.  Other estimates ($> 30$)
had a spread of at least 100 MeV.  The spectra displaying the resonance are
shown in Fig.\ \ref{fig:spec}.  No peak is seen in $\Lambda_c K^- \pi^+$.

We predicted $\tau(\Xi_{cc}^{++,+})$ = (185,53) fs.  A $\Lambda_c K^-
\pi^+$ peak is disfavored by the LHCb lifetime cut $\tau > 150$ fs.  The
$\Xi_{cc}^{++}$ lifetime was measured by LHCb to be $256^{+24}_{-22}\pm14$ fs
\cite{LHCbtau}.  The mass in the $\Xi_c^+ \pi^+$ channel was measured to be
3620.6$\pm$1.5$\pm$0.4$\pm$0.3 MeV \cite{LHCbm}.

The masses of the doubly heavy baryons were calculated with inputs reproducing
the light-quark baryons as shown in Table 2.
\begin{table}[t]
\caption{Masses of ground-state doubly heavy baryons predicted in Ref.\
\cite{PRD90}}
\begin{center}
\begin{tabular}{c c c c} \hline \hline
State & Quark content & $M(J=1/2)$ & $M(J=3/2)$ \\ \hline
$\Xi^{(*)}_{cc}$ & $ccq$ & $3627 \pm 12$ & $3690 \pm 12$ \\
$\Xi^{(*)}_{bc}$ & $b[cq]$ & $6914 \pm 13$ &  $6969 \pm 14$ \\
$\Xi'_{bc}$      & $b(cq)$ & $6933 \pm 12$ &  -- \\
$\Xi^{(*)}_{bb}$ & $bbq$ & $10162 \pm 12$ & $10184 \pm 12$ \\ \hline \hline
\end{tabular}
\end{center}
\end{table}
\begin{table}[h]
\caption{Masses of light-quark baryons predicted with $m_u = m_d \equiv m_q =
 363$ MeV, $m_s=538$ MeV, and hyperfine interaction term $a/(m_q)^2=50$ MeV}
\begin{center}
\begin{tabular}{c c c c} \hline \hline
State (mass     & Spin &   Expression for mass   & Predicted  \\
in MeV)         &     &                               & mass (MeV) \\ \hline
$N(939)$        & 1/2 & $3m_q - 3a/(m_q)^2$       &    939     \\
$\Delta(1232)$  & 3/2 & $3m_q + 3a/(m_q)^2$       &   1239     \\
$\Lambda(1116)$ & 1/2 & $2m_q + m_s - 3a/(m_q)^2$ &   1114     \\
$\Sigma(1193)$  & 1/2 & $2m_q + m_s + a/(m_q)^2-4a/m_q m_s$ & 1179 \\
$\Sigma(1385)$  & 3/2 & $2m_q + m_s + a/(m_q)^2+2a/m_q m_s$ & 1381 \\
$\Xi(1318)$     & 1/2 & $2m_s + m_q + a/(m_s)^2-4a/m_q m_s$ & 1327 \\
$\Xi(1530)$     & 3/2 & $2m_s + m_q + a/(m_s)^2+2a/m_q m_s$ & 1529 \\
$\Omega(1672)$  & 3/2 & $3m_s + 3a/(m_s)^2$       & 1682 \\ \hline \hline
\end{tabular}
\end{center}
\end{table}
One describes light-quark mesons with quark masses $\sim 54$ MeV less.
$M(\Lambda_{c,b}) - M(\Lambda)$ implies $m_{c,b}= (1710.5,5043.5)$ MeV.
These masses are sufficient to describe nonstrange baryons with one
$c$ or $b$ quark, when taking account of deeper $cs$ or $bs$ binding in
baryons with one or two strange quarks and one charm or bottom quark (see
Table 3).  When demanding the same quark masses for mesons and baryons, one
adds 161.5 MeV for a baryon string junction.  The fit quality remains the same.

\begin{table}
\caption{Predicted masses of baryons containing one charm or bottom quark.}
\begin{center}
\begin{tabular}{c c c c c c} \hline \hline
\multicolumn{3}{c}{Charmed baryons} & \multicolumn{3}{c}{Bottom baryons} \\
State ($M$     & Spin & Predicted & State ($M$ & Spin & Predicted \\
in MeV)        &      & $M$ (MeV) & in MeV)    &      & $M$ (MeV) \\ \hline
$\Lambda_c(2286.5)$ & 1/2 & Input &
  $\Lambda_b(5619.5)$ & 1/2 & Input \\
$\Sigma_c(2453.4)$ & 1/2 & 2444.0 &
  $\Sigma_b(5814.3)$ & 1/2 & 5805.1 \\
$\Sigma_c^*(2518.1)$ & 3/2 & 2507.7 &
  $\Sigma_b^*(5833.8)$ & 3/2 & 5826.7 \\
$\Xi_c(2469.3)$ & 1/2 & 2475.3 &
  $\Xi_b(5792.7)$ & 1/2 & 5801.5 \\
$\Xi'_c(2575.8)$ & 1/2 & 2565.4 &
  $\Xi'_b(-)$ & 1/2 & 5921.3 \\
$\Xi^*_c(2645.9)$ & 3/2 & 2628.6 &
  $\Xi^*_b(5949.7)$ & 3/2 & 5944.1 \\
$\Omega_c(2695.2)$ & 1/2 & 2692.1 &
  $\Omega_b(6046.4)$ & 1/2 & 6042.8 \\
$\Omega_c^*(2765.9)$ & 3/2 & 2762.8 &
  $\Omega_b^*(-)$ & 3/2 & 6066.7 \\ \hline  \hline
\end{tabular}
\end{center}
\end{table}

A quark pair is more deeply bound when neither is $u$ or $d$.  For example, the
binding energy of a $c \bar s$ pair is $B(c \bar s) = [3 M(D_s^*) + M(D_s)]/4
- m_s - m_c = - 69.9~{\rm MeV}$.  If one assumes $B(cs)/B(c \bar s) = 1/2$
as for single-gluon exchange then $B(cs) = -35$ MeV.  A similar calculation
gives $B(bs) = -41.8$ MeV.  One must rescale hyperfine interactions when
neither quark is $u$ or $d$.  We take a cue from $M(D_s^*) - M(D_s) \simeq
M(D^*) - M(D)$.

Charm-anticharm binding gives $B(c \bar c) = [3 M(J/\psi) + M(\eta_c)]/4 -
2m_c^m = -258$ MeV, so $B(cc) = -129$ MeV.  Similar calculations give $B(bb) =
-281.4$ MeV and $B(bc) = -167.8 \pm 3.0$ MeV, where the error reflects
uncertainty in the $B^*_c$ mass.  One now can
calculate the doubly heavy ground state baryon masses in Table 1.

A study of isospin splittings in doubly heavy baryons \cite{ispin} was
motivated by the large (60 MeV!) splitting between $\Xi_{cc}^+(3460)$ and
$\Xi_{cc}^{++}(3520)$ claimed by SELEX \cite{SELEX}.  It was found that
$M(\Xi_{cc}^{++})-M(\Xi_{cc}^+) = 2.17 \pm 0.11$ MeV if separate quark masses
are used for light mesons and baryons, or 1.41$\pm 0.12$ MeV if universal
masses are used.  Contributions to mass differences are shown in Table 4.
For details of these calculations and well-obeyed fits to known isosplittings
in light-quark and charmed baryons see Ref.\ \cite{KRi}.  In Table 5 we compare
various predictions for $M(\Xi_{cc}^{++})-M(\Xi_{cc}^+)$.

\begin{table}[h]
\caption{Contributions to isospin splittings (HF=hyperfine interaction) if
(separate, universal) quark masses are used.}
\begin{center}
\begin{tabular}{c c c c} \hline \hline
 Param-  &  Quantity & \multicolumn{2}{c}{Contribution in MeV to} \\
 eter    &           & $M(p) - M(n)$ & $M(\Xi_{cc}^{++}) - M(\Xi_{cc}^+)$ \\
\hline
$\Delta$ & $m_u-m_d$ & --2.48,--2.67 & --2.48,--2.67 \\
  $a$    &  Coulomb  &   1.02,  0.94 &   4.07,  3.77 \\
  $b$    & Strong HF &   0.67,  0.88 & --0.29,--0.33 \\
  $c$    &   EM HF   & --0.51,--0.43 &   0.86,  0.64 \\
         &   Total   & --1.29,--1.29 &   2.17,  1.41  \\ \hline \hline
\end{tabular}
\end{center}
\end{table}

\begin{table}
\caption{Comparison of predictions for isospin splittings of $\Xi_{cc}$ states.}
\begin{center}
\begin{tabular}{c c c} \hline \hline
Author(s)  & Reference & $M(\Xi_{cc}^{++}) - M(\Xi_{cc}^+)$ \\
           &           & (MeV) \\ \hline
Karliner + & PR D {\bf 96}, 033004 (2017)   & $1.41 \pm 0.12^{+0.76}$ \\
Itoh +     & PR D {\bf 61}, 057502 (2000)   & $4.7$ \\
Brodsky +  & PL B {\bf 698}, 251 (2011)     & $1.5 \pm 2.7$ \\
Hwang +    & PR D {\bf 78}, 073013 (2008)   & $2.3 \pm 1.7$ \\
Borsanyi + & Science {\bf 347}, 1452 (2015) & $2.16 \pm 0.11 \pm 0.17$ \\
Lichtenberg& PR D {\bf 16}, 231 (1977)      & $4.7$ \\
Tiwari +   & PR D {\bf 31}, 642 (1985)      & $1.11$ \\
Shah + Rai & EPJC {\bf 77}, 129 (2017)      & --$9$ \\
\hline \hline
\end{tabular}
\end{center}
\end{table}
A spread of values is obtained, but nearly all are at most a few MeV.  Some
authors still entertain the possibility that the SELEX result is correct, with
physics beyond standard model.  This could be put to rest if LHCb sees a
$\Xi_{cc}^+$ at or slightly below 3620 MeV (an observation made more difficult
by its expected short lifetime).

The ``spectator'' process $c \to s W^*$ , where $W^*$ goes to $(e^+\nu_e,
\mu^+\nu_\mu,~3~{\rm colors~of}~u \bar d)$, dominates $\Xi_{cc}^{++}$ decay.
One can emulate kinematic suppression with $x_{cc} \equiv
[M(\Xi_c/M(\Xi_{cc})]^2$:
\begin{center}
$\Gamma(\Xi_{cc}^{++})=\frac{10 G_F^2 M(\Xi_{cc}^{++})^5}{192\pi^3}F(x_{cc})$~,
~~~$F(x) \equiv 1 - 8x + 8x^3 - x^4 +12 \ln(1/x)$,
\end{center}
implying $\tau(\Xi_{cc}^{++}) = 188$~fs.

An additional ``exchange'' process $c d \to s u$ contributes to $\Xi_{cc}^+
= ccd$ decay.  The ``spectator'' partial width is $\Gamma_s = 
\hbar/\tau(\Xi_{cc}^{++}) = \hbar/(256~{\rm fs}) = 2.57 \times 10^{-12}$ GeV,
while the ``exchange'' partial width is $\Gamma_e = 2 [\hbar/\tau(\Xi_c^0) -
\hbar/\tau(\Xi_c^+)] = 5.64 \times 10^{-12}$ GeV.  Here we have used
$\tau(\Xi_c^0) = 154.5 \pm 1.7 \pm 1.6 \pm 1.0$ fs~;~~
$\tau(\Xi_c^+) = 458.8 \pm 3.6 \pm 2.9 \pm 3.1$ fs \cite{LHCblifes}.
Adding the two, $\Gamma_s + \Gamma_e = 8.21 \times 10^{-12}$ GeV
implies $\tau(\Xi_{cc}^+) = 80$ fs, our updated prediction.

One can predict the mass of $\Omega_{cc} = ccs$ using the methods
just described.  The strange quark is about 175 MeV heavier than
nonstrange but more deeply bound to the $cc$ diquark than the nonstrange
quark.  We compare the predictions for $ccq$ and $ccs$ in Table 6.

\begin{table}
\caption{Comparison of predictions for $ccq$ and $ccs$ ground-state baryon
masses.}
\begin{center}
\begin{tabular}{c r c r} \hline \hline
\multicolumn{2}{c}{$\Xi_{cc} = ccq$} & \multicolumn{2}{c}{$\Omega_{cc} = ccs$}
\\
Contribution & Value (MeV) & Contribution & Value (MeV) \\ \hline
$2m_c + m_q$ & 3789.0 & $2m_c + m_s$ & 3959.0 \\
$cc$ binding & ${-}129.0$ & $cc$ binding & ${-}129.0$ \\
$a_{cc}/(m_c)^2$ & 14.2 & $a_{cc}/(m_c)^2$ & 14.2  \\
${-}4a/m_q m_c$ & ${-}42.4$ & ${-}4a'/m_s m_c$ & ${-}42.4$ \\
Total & 3626.8$\phantom{.0}\pm 12$ & Subtotal & 3801.8$\phantom{.0}\pm 12$ \\
\hline \hline
\end{tabular}
\end{center}
\end{table}
The additional binding of $s$ to $cc$ is $-109.4 \pm 10.5$ MeV, giving
$M(\Omega_{cc}) = 3692 \pm 16$ MeV, $M(\Omega_{cc}^*) = 3756 \pm 16$ MeV,
With universal quark masses and a 161.5 MeV ``string junction'' term for
baryons one predicts $M(\Omega_{cc}) \sim 40$ MeV higher.

M. Karliner and I investigated $QQ' \bar u
\bar d$ systems \cite{bbud}, where $Q,Q' = c$ or $b$.  We found $c c \bar u
\bar d$ unbound; it could decay to $DD^*$ or $D D \gamma$.  The lowest-lying $b
c \bar u \bar d$ state was near $B D \gamma$ threshold and could be
bound.  We predicted $M(b b \bar u \bar d)
= 10,389 \pm 12$ MeV, 215 MeV below $B^- B^{*0}$ threshold and 170 MeV below
$B^- B^0 \gamma$ threshold.  Regarding $b b$ as a color-$3^*$ diquark
(transforming under QCD as an antiquark), fermi statistics required its spin to
be 1.  The lightest $\bar q \bar q'$ state ($q,q' = u,d$) is a color-3 $\bar u
\bar d$ state with isospin zero; fermi statistics require its spin to be zero.
The mass prediction then relies on accounting for constituent-quark masses,
hyperfine interactions, and binding effects (Table 7).

\begin{table}
\caption{Contributions (in MeV) to mass of lightest $Q Q' \bar q \bar q'$
tetraquark.}
\begin{center}
\begin{tabular}{c r c r c r} \hline \hline
\multicolumn{2}{c}{$c c\bar u \bar d,~J^P = 1^+$} &
\multicolumn{2}{c}{$b c\bar u \bar d,~J^P = 0^+$} &
\multicolumn{2}{c}{$b b\bar u \bar d,~J^P = 1^+$} \\ \hline
Contribution & Value & Contribution & Value & Contribution & Value \\ \hline
   $2m_c^b$ & 3421.0 & $m_b + m_c$ & 6754.0 &   $2m^b_b$ & 10087.0 \\
   $2m^b_q$ &  726.0 &    $2m^b_q$ &  726.0 &   $2m^b_q$ &  726.0  \\
$cc$ hyperfine & 14.2 & $bc$ hyperfine & $-25.5$ & $bb$ hyperfine & 7.8 \\
${-}3a/(m^b_q)^2$ & ${-}150.0$ & ${-}3a/(m^b_q)^2$ & ${-}150.0$ &
  ${-}3a/(m^b_q)^2$ & ${-}150.0$ \\
$cc$ binding & ${-}129.0$ & $bc$ binding & ${-}170.8$ &
  $bb$ binding & ${-}281.4$ \\
Total & 3882$\pm$12 & Total & 7134$\pm$13 &
  Total & $10389\pm12$ \\ \hline \hline
\end{tabular}
\end{center}
\end{table}
Spin zero is allowed for the $bc \bar u \bar d$ state, taking advantage of the
attractive $b c$ hyperfine interaction.  Since $M(cc \bar u \bar d) > M(D^0) +
M(D^+) = 3734$ MeV, it can decay to $D^0 D^+ \gamma$ (decay to $D^0 D^+$ is
forbidden).  We cannot tell whether $M(b c \bar u \bar d)$ is less than
$M(D^0) + \bar M(B^0) = 7144$ MeV.  The estimated lifetime of the $bb \bar u
\bar d$ state is 367 fs.

The LHCb Collaboration has presented evidence for five narrow $\Omega_c^*$
states decaying to $\Xi_c^+ K^-$ \cite{LHCbOm}.  (Already known were the
ground $css$ states:  $\Omega_c(2695,1/2^+)$ and $\Omega_c^*(2766,3/2^+)$
\cite{PDG}.)  Marek Karliner and I \cite{KROm} identified the narrow states as
five P-wave excitations, with an alternative assignment of the two highest-mass
states as positive-parity radial excitations of the ground states.  In that
case two $J^P = 1/2^-$ states would remain to be seen, one around 2904 MeV
decaying to $\Omega_c \gamma$ and/or $\Omega_c \pi^0$, and the other around
2978 MeV decaying to $\Xi_c^+ K^-$ in an S-wave.

What does it cost to excite a hadron from S-wave to P-wave \cite{KRP}?
Defining a residual energy $\Delta E_R \equiv \Delta E_{PS}$--%
$B_{12}$, where $B_{12}$ is the binding energies of constituents, we found a
good fit with $\Delta E_R = (417.37 - 0.2141 \mu_{12})$ MeV, where
$\mu_{12}$ is the reduced mass.

The prospects for exotic mesons and baryons (beyond $q \bar q$ and $qqq$)
are bright.  They {\it do} exist; molecular configurations are at least part
of the story.  Heavy quarks have a lower kinetic energy and help to stabilize
exotic configurations containing them.  Techniques for mass estimation
(constituent-quark masses, hyperfine interactions, and binding effects) are
relatively straightforward and are starting to be tested for $Q Q' q$ baryons.
One frontier is states $Q_1 Q_2 \bar Q_3 \bar Q_4$ with all quarks heavy.  Are
there any $c c \bar c \bar c$ lighter than $2 M(\eta_c)$?  Are there any $b b
\bar b \bar b$ lighter than $2 M(\eta_b)$?  Can the quark-level analogue of
nuclear fusion \cite{Fusion} be put to use?  Still to be known is what it costs
to produce one or more extra heavy quarks via the strong interactions.
When do two heavy quarks end up in the same hadron?

My thanks to M. Gronau and M. Karliner for many enjoyable collaborations, and
to the Organizers and to the Enrico Fermi Institute for their welcome support.

\end{document}